\documentclass{aastex63}
\usepackage{bm}
\received{\today}
\revised{\today}
\accepted{\today}
\submitjournal{ApJL}

\shorttitle{Exomoon Constraints}
\shortauthors{Quarles, Li, \& Rosario-Franco}
\begin{document}

\title{Application of Orbital Stability and Tidal Migration Constraints for Exomoon Candidates}

\correspondingauthor{Billy Quarles}
\email{billylquarles@gmail.com}

\author[0000-0002-9644-8330]{Billy Quarles}
\affil{Center for Relativistic Astrophysics, School of Physics, 
Georgia Institute of Technology, Atlanta, GA 30332 USA}

\author[0000-0001-8308-0808]{Gongjie Li}
\affil{Center for Relativistic Astrophysics, School of Physics, 
Georgia Institute of Technology, Atlanta, GA 30332 USA}

\author[0000-0003-0216-559X]{Marialis Rosario-Franco}
\affiliation {National Radio Astronomy Observatory, Socorro NM 87801, USA}
\affiliation{University of Texas at Arlington, Department of Physics, Arlington TX 76019, USA}



\begin{abstract}
Satellites of extrasolar planets, or exomoons, are on the frontier of detectability using current technologies and theoretical constraints should be considered in their search.  In this Letter, we apply theoretical constraints of orbital stability and tidal migration to the six candidate KOI systems proposed by \cite{Fox2020} to identify whether these systems can potentially host exomoons.  The host planets orbit close to their respective stars and the orbital stability extent of exomoons is limited to only $\sim$40\% of the host planet’s Hill radius ($\sim$20 R$_{\rm p}$).  Using plausible tidal parameters from the solar system, we find that four out of six systems would either tidally disrupt their exomoons or lose them to outward migration within the system lifetimes.  The remaining two systems (KOI 268.01 and KOI 1888.01) could host exomoons that are within 25 R$_{\rm p}$ and less than $\sim$3\% of the host planet’s mass.  However, a recent independent transit timing analysis by \cite{Kipping2020b} found that these systems fail rigorous statistical tests to validate them as candidates.  Overall, we find the presence of exomoons in these systems that are large enough for TTV signatures to be unlikely given the combined constraints of observational modeling, tidal migration, and orbital stability.  {Software to reproduce our results is available in the GitHub repository: Multiversario/satcand}.

\end{abstract}



\section{Introduction} \label{sec:intro}
The \textit{Kepler} data has discovered a myriad of exoplanets, however a substantial number of viable planet satellite (exomoon) candidates have not been uncovered.  The best exomoon candidate (Kepler 1625b-I, \cite{Teachey2018b}) is hosted by a Jupiter-sized exoplanet on a fairly wide orbit ($\sim$287 days).  \cite{Fox2020} recently identified six KOIs (\text{Kepler} Objects of Interest) that exhibit transit timing variations \citep[TTVs, ][]{Kipping2009a,Kipping2009b} which could possibly be explained by the reflex motion of an exomoon.  If validated, such a discovery would represent a giant leap forward in the detection of exomoons \citep{Kipping2012,Kipping2013a,Kipping2013b,Kipping2014,Kipping2015b,Teachey2018a}.  A major difference between these KOIs and Kepler-1625b is the proximity to their host star, where gravitational tides and/or general relativity effects can be important.  We provide an analysis focusing on the orbital stability limits for exomoons \citep{Rosario-Franco2020} and the possible outcomes of tidal migration considering the tidal influence between the planet-star and planet-satellite \citep{Sasaki2012}.

The search for exomoons using photometric data \citep{Sartoretti1999,Cabrera2007} now has a long history due to the \textit{Kepler} mission, where additional constraints beyond TTVs are usually required (e.g., transit duration variations, or TDVs, \cite{Kipping2009a}), or techniques that make use of sampling effects \cite{Heller2014,Hippke2015,Heller2016}.  \cite{Kipping2020b} performed an independent analysis of the KOIs proposed by \cite{Fox2020} and found no compelling for evidence among the six candidates using rigorous statistical hypothesis testing.  Kepler-1625b passes 2 out of 3 such tests and remains the best exomoon candidate despite its own history \citep{Heller2018,Heller2019,Kreidberg2019}.  \cite{Kipping2020a} have introduced constraints from tidal interactions \citep{Barnes2002} that place limits on allowable ranges from TTVs or TDVs, however tidal interactions that change the planetary rotation also need to be included because of the non-negligible effect on the moon lifetimes \citep[][see their Figure 13]{Sasaki2012}.

Gravitational tidal models depend on parameters (e.g., tidal Love number $k_{2}$, tidal time lag $\Delta t$, moment of inertia $\alpha$, or tidal quality factor $Q$) that are unconstrained for most (if not all) exoplanets and even not well constrained for planets in our own solar system \citep{Goldreich1966,Lainey2016}.  Models based upon equilibrium tides with a constant time lag \citep{Hut1981,Eggleton1998,Fabrycky2007} or  with a constant $Q$ \citep{Goldreich1966,Ward1973} are qualitatively similar in their predictions of moon lifetimes \citep{Tokadjian2020}, where discrepancies may arise long after the main sequence lifetime of the host stars.  Although these parameters are not well known for exoplanets, the tidal migration largely depends on the ratio $k_2/(\alpha Q)$ and reasonable extremes can be estimated from the solar system planets.

In this Letter, we determine the plausibility of exomoons orbiting the six candidates from \cite{Fox2020} using orbital stability \citep{Rosario-Franco2020}, a constant Q tide model \citep{Sasaki2012}, and results from a recent TTV analysis \citep{Kipping2009b}.  In Section \ref{sec:orb_stab}, we demonstrate how orbital stability limits can be used to place upper limits on physical parameters of exomoons.  We evaluate a constant Q tide model and estimate the lifetime of exomoons in Section \ref{sec:tide_mig}.  We combine our analysis of exomoon orbital stability and tidal migrations with the upper limits from \cite{Kipping2020b} in Section \ref{sec:upper}.  Our results are summarized in Section \ref{sec:conc}, where we also identify how Kepler 1625b-I fits within our analysis.

\section{Orbital Stability} \label{sec:orb_stab}
An exomoon gravitationally interacts with both its host planet and the planet's host star, where the combination of these forces limits the orbital separation between the exomoon and its host planet.  The limiting planet-satellite separation, or stability limit, is a fraction $f_{\rm crit}$ of the the Hill radius R$_H$ (=$a_{\rm p}$[(M$_{\rm p}$+M$_{\rm sat}$)/(3M$_\star$)]$^{1/3}$), which depends  on the planetary semimajor axis $a_{\rm p}$, planetary mass M$_{\rm p}$, satellite mass M$_{\rm sat}$, and the stellar mass M$_\star$.  Our recent work \citep{Rosario-Franco2020} identified $f_{\rm crit} \approx 0.4061$ through a large number of N-body simulations that varied the initial planet-satellite separation $a_{\rm sat}$, planet eccentricity $e_{\rm p}$ and satellite mean anomaly $MA_{\rm sat}$.  We define the stability limit as: $a_{\rm crit} = f_{\rm crit}R_H(1-1.1257e_{\rm p})$ in terms of the Hill radius, where the additional factor is necessary to account for changes in the Hill radius for eccentric orbits of the planet.  

Although the planetary semimajor axis is well-determined, there is a significant uncertainty in the stellar mass for the six exomoon candidate systems proposed by \cite{Fox2020}.  Moreover, the planetary mass is undetermined and we must rely on probabilistic determinations \citep{Chen2017} based upon statistical relationships uncovered from the confirmed \textit{Kepler} planets with radial velocity mass measurements.  We summarize the current values and uncertainties obtained from the \textit{Kepler} Exoplanet Archive (DR25) for the stellar mass M$_\star$, planetary radius R$_{\rm p}$, planetary semimajor axis $a_{\rm p}$, and system age $\tau$ in Table \ref{tab:KOI_params}.  Updated values are used based upon studies that implement asteroseismology  \citep{SilvaAguirre2015} or better isochrone fitting \citep{Morton2016} for the stellar age.  \cite{Berger2018} identifies better constraints on the planet radius R$_{\rm p}$ due to precise astrometric measurements from Gaia, where we update appropriately.  The planetary mass is estimated using \texttt{Forecaster} from \cite{Chen2017} based upon our best knowledge of the planet radius and the satellite mass is small compared to the planetary mass.

\begin{figure}
    \centering
    \includegraphics[width=\linewidth]{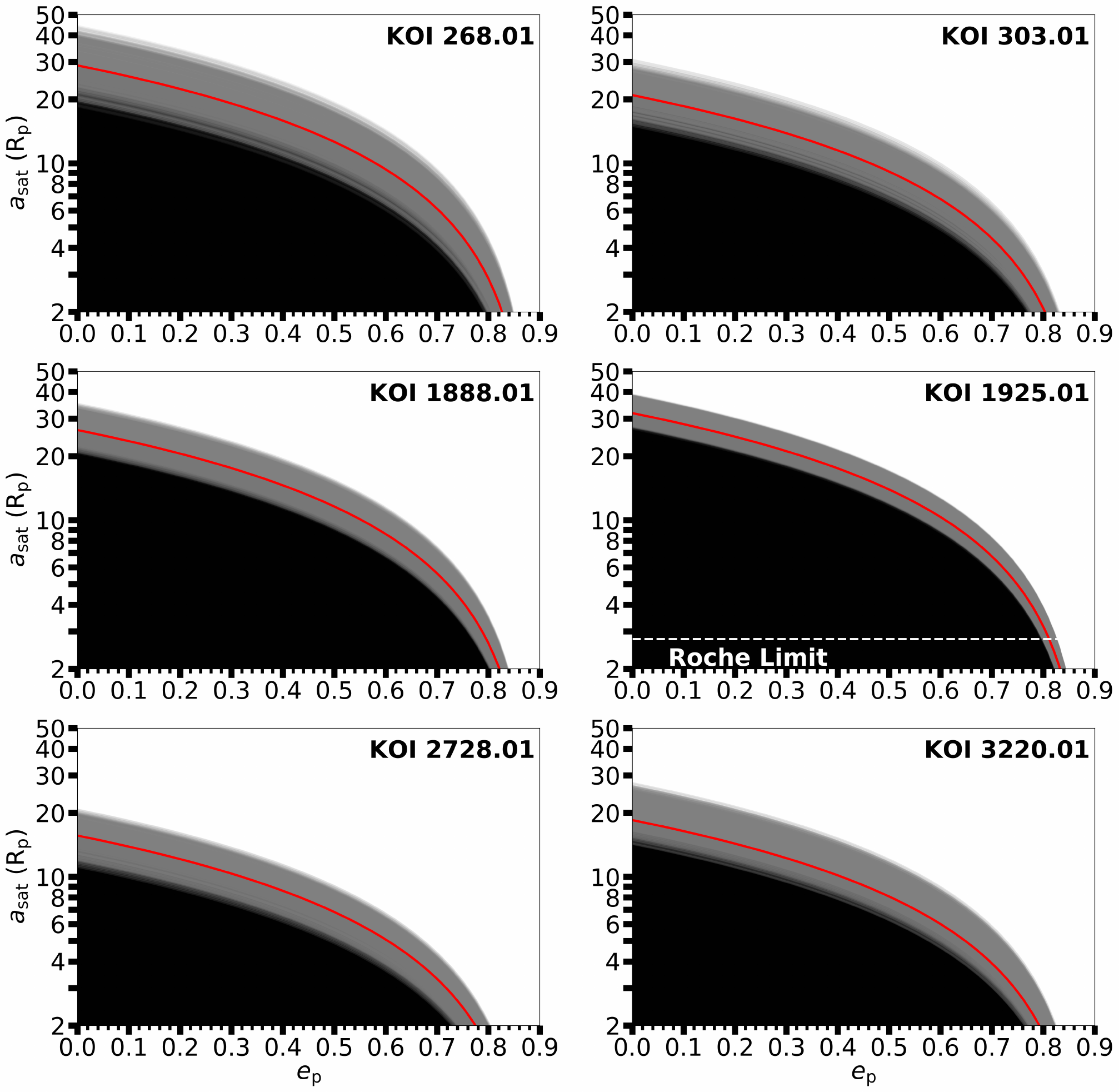}
    \caption{The range in exomoon semimajor axis $a_{\rm sat}$ for each of the six \textit{Kepler} KOIs proposed by \cite{Fox2020} is constrained using our updated outer stability limit formula \citep{Rosario-Franco2020} as a function of the planetary radius $R_{\rm p}$, where the black region marks the stable exomoon regime as a function of assumed planetary eccentricity and the white region denotes parameters that are quickly lost due to gravitational perturbations.  The red curve shows the outer stability limit using the mean parameters for each system (see Table \ref{tab:KOI_params}) and the gray curves indicate how the outer limit changes in response to observational or modeling uncertainties.  {The estimated Roche limit for most of the KOI candidates is below 2 R$_{\rm p}$, except for KOI 1925.01, where its Roche limit is marked with a horizontal dashed white line.}} 
    \label{fig:exomoon_stab}
\end{figure}

Using our formalism for the stability limit and the best known system parameters (Table \ref{tab:KOI_params}), we identify the location of $a_{\rm crit}$ in units of the planetary radius R$_{\rm p}$ and as a function of the planetary eccentricity in Figure \ref{fig:exomoon_stab}.  The red curve marks the determination of the stability limit using the mean system values and the gray curves illustrate the variance in the stability limit due to the uncertainties in the system values.  The black region denotes the  combinations of satellite semimajor axis $a_{\rm sat}$ and planet eccentricity $e_{\rm p}$ that permit long-term stability.  We use a lower boundary on $a_{\rm sat} = 2$ R$_{\rm p}$, but the lower boundary should be defined by the Roche limit.  The Roche limit depends on unknown properties (mass or density) of the exomoon candidates and their host planets.  {Using the mean values of the probabilistic planetary masses, we can estimate some sensible values for the Roche limit.  The Roche limit for KOI 1925.01 is $\sim2.75$ R$_{\rm p}$, while the Roche limit for all the other KOIs is less than 2 R$_{\rm p}$.}  Despite the unknowns, we can estimate the stability limit $a_{\rm crit}$ within a factor of $\sim$2.  \cite{Kipping2020b} identified a large eccentricity ($e_{\rm p} \sim 0.6$) for KOI 1925.01 through his photodynamical fits, which substantially truncates the stability limit for exomoons in the system so that the largest planet-satellite separation is $a_{\rm sat} \lesssim 8-12$ R$_{\rm p}$.

\section{Tidal Migration}\label{sec:tide_mig}
Tidal migration timescales and/or distances can be used to constrain the possibility of an exoplanet to host exomoons \citep{Barnes2002,Sucerquia2019}.  The migration depends on several parameters that are unknown (tidal Love number $k_{2p}$ and tidal Quality factor $Q_{\rm p}$), but we can identify plausible parameters using values from the solar system.  Using the observed planetary radius R$_{\rm p}$, we assign either 0.299 (R$_{\rm p} < 2$ R$_\oplus$; \cite{Lainey2016}) or 0.12 (R$_{\rm p} \geq 2$ R$_\oplus$; \cite{Gavrilov1977}) for the tidal Love numbers.  A lower limit for $Q_{\rm p}$ can be estimated using the system age $\tau$ and the critical mean motion $n_{\rm crit}$ (=$\sqrt{G(M_{\rm p}+M_{\rm sat})/a_{\rm crit}^3}$) determined from the stability limit $a_{\rm crit}$.  We parameterize the planet-satellite mass ratio as $f_m = M_{\rm sat}/M_{\rm p}$ and evaluate tidal models over a wide range ($10^{-3} \leq f_m \leq 10^{-1}$). 

We implement a constant Q tidal model \citep{Sasaki2012} that is directly applicable to planet-satellite mass ratios $M_{\rm sat}/M_{\rm p} < 0.1$, which is akin to the Pluto-Charon system \citep{Cheng2014}.  Through our tidal model, we are interested in two regimes: 1) the satellite tidally migrates outward past the stability limit (see \S\ref{sec:orb_stab}) before the satellite's mean motion synchronizes with the planetary spin frequency ($\Omega_{\rm p} = n_{\rm sat}$) or 2) the satellite tidally migrates inward towards the Roche limit following angular momentum conservation after synchronization.  \cite{Sasaki2012} provides an analytical decision tree algorithm that is based on the following differential equations:

\begin{eqnarray}
    \dot{n}_{\rm sat} &=& -\frac{9}{2} \frac{k_{2p}R_{\rm p}^5}{Q_{\rm p}} \frac{M_{\rm sat}}{M_{\rm p}} \frac{n_{\rm sat}^{16/3}}{[G(M_{\rm p}+M_{\rm sat})]^{5/3}} {\rm sgn}[\Omega_{\rm p} - n_{\rm sat}], \label{eqn:n_sat} \\
    \dot{n}_{\rm p} &=& -\frac{9}{2} \frac{k_{2p}R_{\rm p}^5}{Q_{\rm p}} \frac{n_{\rm p}^{16/3}}{G(M_{\rm p}+M_{\rm sat})[G(M_\star+M_{\rm p}+M_{\rm sat})]^{5/3}} {\rm sgn}[\Omega_{\rm p} - n_{\rm p}], \label{eqn:n_p}\\
    \dot{\Omega}_{\rm p} &=& -\frac{3}{2}\frac{k_{2p}R_{\rm p}^3}{\alpha Q_{\rm p}}\left[\frac{GM_{\rm sat}^2}{[GM_{\rm p}]^3}n_{\rm sat}^4 {\rm sgn}[\Omega_{\rm p} - n_{\rm sat}] + \frac{n_{\rm p}^4}{GM_{\rm p}}{\rm sgn}[\Omega_{\rm p} - n_{\rm p}]\right], \label{eqn:Omg_p}
\end{eqnarray}
which depends on the exomoon's mass M$_{\rm sat}$, planetary mean motion $n_{\rm p}$, and the moment of inertia constant $\alpha$.  Equations 1--3 are valid assuming that the exomoon's orbit is not yet synchronized with the planetary rotation ($\Omega_{\rm p} > n_{\rm sat}$), the exomoon spin $\Omega_{\rm sat}$ synchronous with its mean motion ($\Omega_{\rm sat}=n_{\rm sat}$), and the planetary spin is large compared to its mean motion ($\Omega_{\rm p} > n_{\rm p}$).  Moreover, these equations are applicable for circular and coplanar orbits.  Eccentric planetary orbits are beyond our scope because only one of the candidates has an estimate for the planetary eccentricity, but these equations can be modified by including a polynomial function $N(e)$ \citep[e.g.,][]{Cheng2014}.  

After synchronization between the satellite mean motion and planetary rotation ($\Omega_{\rm p} = n_{\rm sat}$), the planet-satellite system evolves through angular momentum conservation.  {The total angular momentum $L$ consists of the sum of three terms: 1) the planetary rotational angular momentum, 2) the planetary orbital angular momentum, and 3) the satellite orbital angular momentum, which is represented by:}

\begin{equation}
    {L = \alpha M_p R_p^2\Omega_p + \frac{M_p[G(M_\star+M_p+M_{sat})]^{2/3}}{n_p^{1/3}} + \frac{\mu [G(M_p+M_{sat})]^{2/3}}{n_{sat}^{1/3}},}
\end{equation}

{which includes the reduced mass $\mu = (M_p M_{sat})/(M_p + M_{sat})$.  Substituting $\Omega_p = n_{sat}$ and taking the first derivative $\dot{L}$, we obtain the differential equations that evolve due to angular momentum conservation as:}
\begin{equation} \label{eqn:n_sat_sync}
    \dot{n}_{\rm sat} = -\frac{M_{\rm p}[G(M_\star+M_{\rm p}+M_{\rm sat})]^{2/3}n_{\rm p}^{-4/3}\dot{n}_{\rm p}}{{\mu}[G(M_{\rm p}+M_{\rm sat})]^{2/3}n_{\rm sat}^{-4/3}-3\alpha R_{\rm p}^2 M_{\rm p}},
\end{equation}
the argument for the sgn function in Equation \ref{eqn:n_p} is replaced with [$n_{\rm sat}-n_{\rm p}$], and the planetary rotation follows the satellite mean motion evolution ($\dot{\Omega}_{\rm p}=\dot{n}_{\rm sat}$), which spins up the planet as the satellite spirals inward.  Equation \ref{eqn:n_sat_sync} is modified from Equation 14b in \cite{Sasaki2012} to include all of the masses, including a reduced-mass factor ${\mu}$ on the exomoon's orbital angular momentum \citep{Cheng2014}.

\begin{figure}
    \centering
    \includegraphics[width=\linewidth]{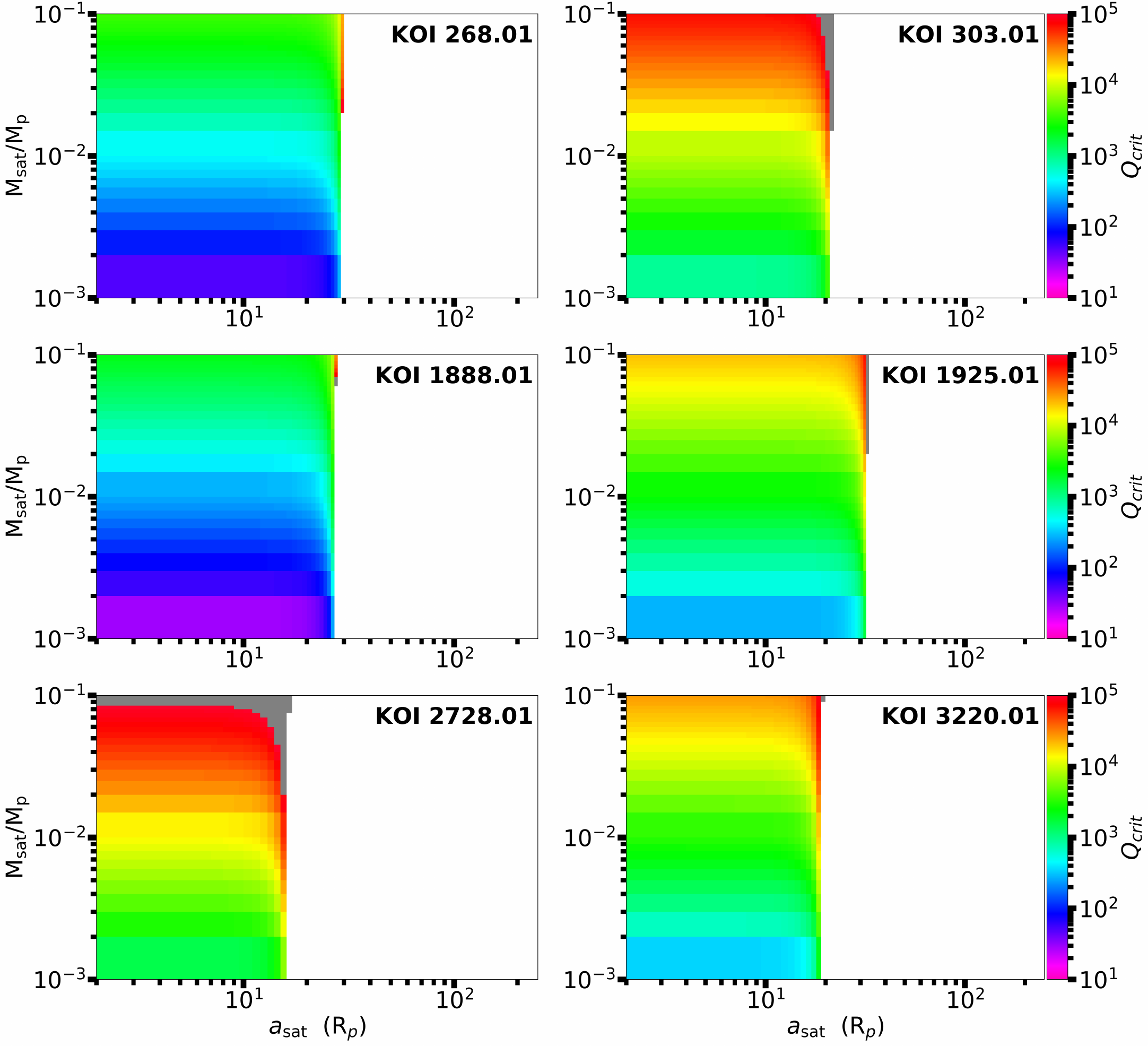}
    \caption{The minimum planetary tidal quality factor $Q_{\rm crit}$ (color-coded) that allows for an exomoon to survive beyond the current system lifetime $\tau$ for each of the six candidate KOIs.  The mean values are used for the stellar mass, planetary radius, and planetary mass from Table \ref{tab:KOI_params}, where $k_{2p} = 0.299$ for Earth-like planets \citep{Lainey2016} for KOI 1925.01 and $k_{2p}=0.12$ \citep{Gavrilov1977} for all the other Neptune-like candidates.  The white region denotes that the exomoon separation has exceeds the outer stability limit within the system lifetime {and the gray region marks when} $Q_{\rm crit}> 10^5$, which is unrealistic given our knowledge of the solar system giant planets. } 
    \label{fig:exomoon_tides}
\end{figure}

Conditions for regime (1) can be determined by first integrating Equation \ref{eqn:n_sat} analytically and setting the result equal to the critical mean motion $n_{\rm crit}$.  The tidal quality factor $Q_{\rm p}$ is proportional to the total tidal migration timescale $T$, where $Q_{\rm p}$ has to be sufficiently large so that the exomoon can begin at a given $a_{\rm sat}$ and remain bound for at least the system age $\tau$.  A similar approach is used by \citeauthor{Barnes2002} to prescribe limits for the satellite mass {\citep[][see their Equation 8]{Barnes2002}}, where we solve for $Q_{\rm p}$ instead.  As a result, we obtain a lower limit for $Q_{\rm p}$ as:

\begin{equation} \label{eqn:Q_crit}
    Q_{\rm crit} \geq \frac{39}{2} \frac{k_{2p}R_{\rm p}^5\tau M_{\rm sat}{\sqrt{G(M_{\rm p}+M_{\rm sat})}}}{{M_{\rm p}}\left(a_{\rm crit}^{13/2}-a_{\rm sat}^{13/2}\right)},
\end{equation}

where a tidal quality factor below the critical value ($Q_{\rm p} < Q_{\rm crit}$) will migrate outward past the stability limit on a timescale less than the system lifetime $\tau$. Figure \ref{fig:exomoon_tides} shows this lower limit $Q_{\rm crit}$ (color-coded; log scale) for each of the six exomoon candidate systems as a function of the planet-satellite mass ratio $M_{\rm sat}/M_{\rm p}$ and initial separation $a_{\rm sat}$ on a logarithmic scale.  Tidally unstable conditions are colored white {and unrealistic conditions $Q_{\rm crit}>10^5$ are colored gray}.  The lower limit $Q_{\rm crit}$ is evaluated using the mean values from Table \ref{tab:KOI_params}, where the observational uncertainties in the planetary radius, planetary mass, and the system age shift these values slightly.  Equation \ref{eqn:Q_crit} shows that uncertainties in the planetary radius drive the largest changes and it is one of the better constrained observational quantities.  

We can also infer a plausible value for $Q_{\rm p}$ from the planetary radius as long as the host planet is not in the an ambiguous region \citep{Rogers2015,Chen2017}.  KOI 1925.01 is nearly Earth-sized, where we can estimate that its $Q_{\rm p}\lesssim 200$ and regions with $Q_{\rm crit}\gtrsim 200$ could be excluded ({light blue to red}).  This is justified because all of the terrestrial planets in the solar system have $Q_{\rm p} \lesssim 100$ and specifically for the Earth $Q_{\rm p} \approx 12$ \citep{Lainey2016}.  A similar approach can be applied to the other KOIs using a very uncertain estimate for Neptune's $Q_{\rm p} \sim 1000$ \citep{Lainey2016}, thereby excluding regions with $Q_{\rm crit} \gtrsim 2000$ (light green to red).  These conditions place constraints on KOI 303.01, KOI 1925.01, KOI 2728.01, and KOI 3220.01 to allow for exomoons that are less than 1\% of the planetary mass.

\begin{figure}
    \centering
    \includegraphics[width=\linewidth]{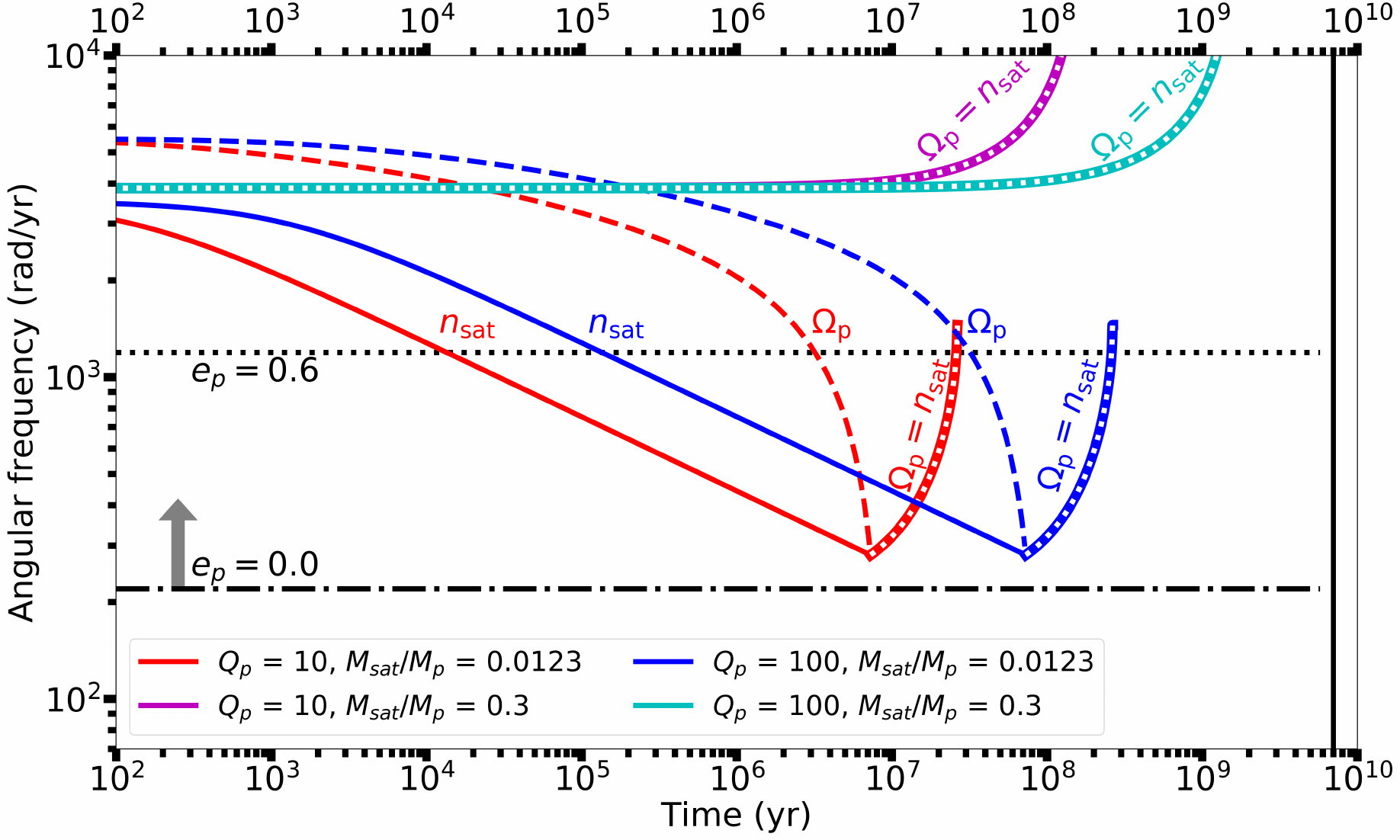}
    \caption{Evolution using the mean parameters from KOI 1925.01 for a putative satellite's mean motion $n_m$ (solid) and the planet's spin frequency $\Omega_{\rm p}$ (dashed) using a constant Q tidal model \citep{Sasaki2012}, where the initial satellite separation is 5 $R_{\rm p}$ and the planetary rotation period begins at 10 hours. The mean values are used for the stellar mass, planetary radius, and planetary mass from Table \ref{tab:KOI_params}, where a vertical solid (black) line marks the mean system lifetime $\tau$ and a horizontal (dash-dot) line denotes the critical mean motion $n_{\rm crit}$ corresponding to the outer stability limit \citep{Rosario-Franco2020}.  The satellite's mean motion and planetary rotation synchronize ($\Omega_{\rm p} = n_m$) causing the solid and dashed curves to overlap (solid with white dots).  For the high mass ratio case ($M_{\rm sat}/M_{\rm p} = 0.3$), the synchronization occurs rapidly.  During inward migration, the slope of the satellite's mean motion rapidly increases and marks the impending collision with the planet.   }  
    \label{fig:KOI1925_tides}
\end{figure}

The initial values for the planetary spin frequency must be much larger than the satellite's mean motion ($\Omega_{\rm p}>> n_{\rm sat}$) for the above conditions to hold, which is the case considering an initial $\Omega_{\rm p}$ near break-up.  For slower planetary rotation rates, we must consider the planet-satellite system evolution using angular momentum conservation (Equation \ref{eqn:n_sat_sync}) and evaluate whether the infall timescale is less than the system age $\tau$.  Figure \ref{fig:KOI1925_tides} illustrates a numerical solution of KOI 1925.01 using Equations \ref{eqn:n_sat}--\ref{eqn:Omg_p} ($\Omega_{\rm p} > n_{\rm sat}$) or Equation \ref{eqn:n_sat_sync} with a modified Equation \ref{eqn:n_p} ($\Omega_{\rm p} = n_{\rm sat}$) using a Runge-Kutta-Fehlberg integration scheme\footnote{{A repository is available on\dataset[GitHub]{https://github.com/Multiversario/satcand} (and archived on\dataset[Zenodo]{https://10.5281/zenodo.4026288}) containing python scripts that reproduce our results and figures.}} \citep[\texttt{scipy};][]{2020SciPy_NMeth} with an absolute and relative tolerance of $10^{-12}$.  The time evolution of $\Omega_{\rm p}$ and $n_{\rm sat}$ are evaluated assuming that the host planet is Earth-like in its tidal Love number ($k_{2p} =0.299$), the initial rotation period is 10 hours, the initial planet-satellite separation is 5 R$_{\rm p}$, and we use the mean values for the stellar mass, planetary radius, and system age.  We evaluate two values in $Q_{\rm p}$ (10 and 100), as well as two mass ratios (0.0123 and 0.3) that are color-coded in the legend.  For the Earth-Moon mass ratio ($M_{\rm sat}/M_{\rm p} = 0.0123$) case, the planetary spin (dashed) evolves following Equation \ref{eqn:Omg_p} and the satellite mean motion (solid) evolves following Equation \ref{eqn:n_sat} until $\Omega_{\rm p} = n_{\rm sat}$ and follows Equation \ref{eqn:n_sat_sync} once synchronized.  The planetary spin angular momentum is insufficient to drive the satellite past the stability limit for a circular orbit (horizontal dash-dot line), but the infall phase ultimately destroys the satellite.  The timescale for this evolution increases linearly with the assumed $Q_{\rm p}$ and a $Q_{\rm p}$ that is much larger than terrestrial values is necessary to prolong the satellite lifetime enough to be observed by \textit{Kepler}.  Moreover, if we use the truncated stability limit assume $e_{\rm p} = 0.6$ (horizontal dotted line), then the satellite can be stripped away within $\sim$10$^5$ years.  

As the planet-satellite mass ratio increases, the satellite mean motion synchronizes with the host planet spin rapidly and nearly all of the evolution follows angular momentum conservation (Equation \ref{eqn:n_sat_sync}).  \cite{Cheng2014} showed a similar evolution with the Pluto-Charon system, where Pluto's tidal Love number ($k_{2p} = 0.058$) is significantly smaller than the terrestrial planets.  Using KOI 1925.01 with a larger mass ratio ($M_{\rm sat}/M_{\rm p} = 0.3$), Figure \ref{fig:KOI1925_tides} shows the satellite mean motion evolution to remain steady for the first 10$^7$ years, but eventually enters an inspiral phase, where a larger $Q_{\rm p}$ delays the demise proportionally ($\dot{n}_{\rm sat} \propto (n_{\rm sat}/n_{\rm p})^{4/3}\dot{n}_{\rm p}$).  To prolong the satellite lifetime to equal the system lifetime, a large dissipation factor is needed ($Q_{\rm p} \sim 700$) and is unrealistic compared with the terrestrial planets.  

\section{Combining Limits from Observational Modeling, Orbital Stability, and Tidal Migration}\label{sec:upper}
Analysis of the \textit{Kepler} data can uncover the planetary radius, planetary orbital period, and even estimates for the stellar mass and age using asteroseismology \citep{SilvaAguirre2015}. \cite{Fox2020} used transit timing variations (TTVs) to suggest an unseen perturber within six KOI systems, which could be caused by gravitational interactions with an exomoon.  Additionally, \cite{Fox2020} prescribe a 1 R$_\oplus$ transit depth threshold for the proposed satellite because it otherwise would have been detected in the \textit{Kepler} data.  This puts an upper limit on the mass ratio to $\sim$0.1--0.3 for 5 of 6 KOI candidates, where KOI 1925.01 could be significantly higher ($M_{\rm sat}/M_{\rm p} \lesssim 0.8$).  However, high mass ratio planets would produce identifiable distortions (blended or w-shaped transits; \cite{Lewis2015}) to the light curve.  We adjust this threshold lower to 0.5 R$_\oplus$ because such distortions are not apparent in the light curves presented in \cite{Kipping2020b} and assume a Mars-like density to derive the respective satellite mass.  Additionally, there is a threshold set by the TTV amplitude and we adopt the 3$\sigma$ constraints shown in \cite{Kipping2020b}. From Section \ref{sec:orb_stab}, we apply an orbital stability constraint \citep{Rosario-Franco2020} assuming a circular planetary orbit.  In Section \ref{sec:tide_mig}, we introduce constraints based upon tidal migration \citep{Sasaki2012,Cheng2014}, where bound exomoons are possible for $Q_{\rm crit}\lesssim 2000$ (Neptune-like) or $Q_{\rm crit} \lesssim 200$ (Earth-like) host planets.

\begin{figure}
    \centering
    \includegraphics[width=\linewidth]{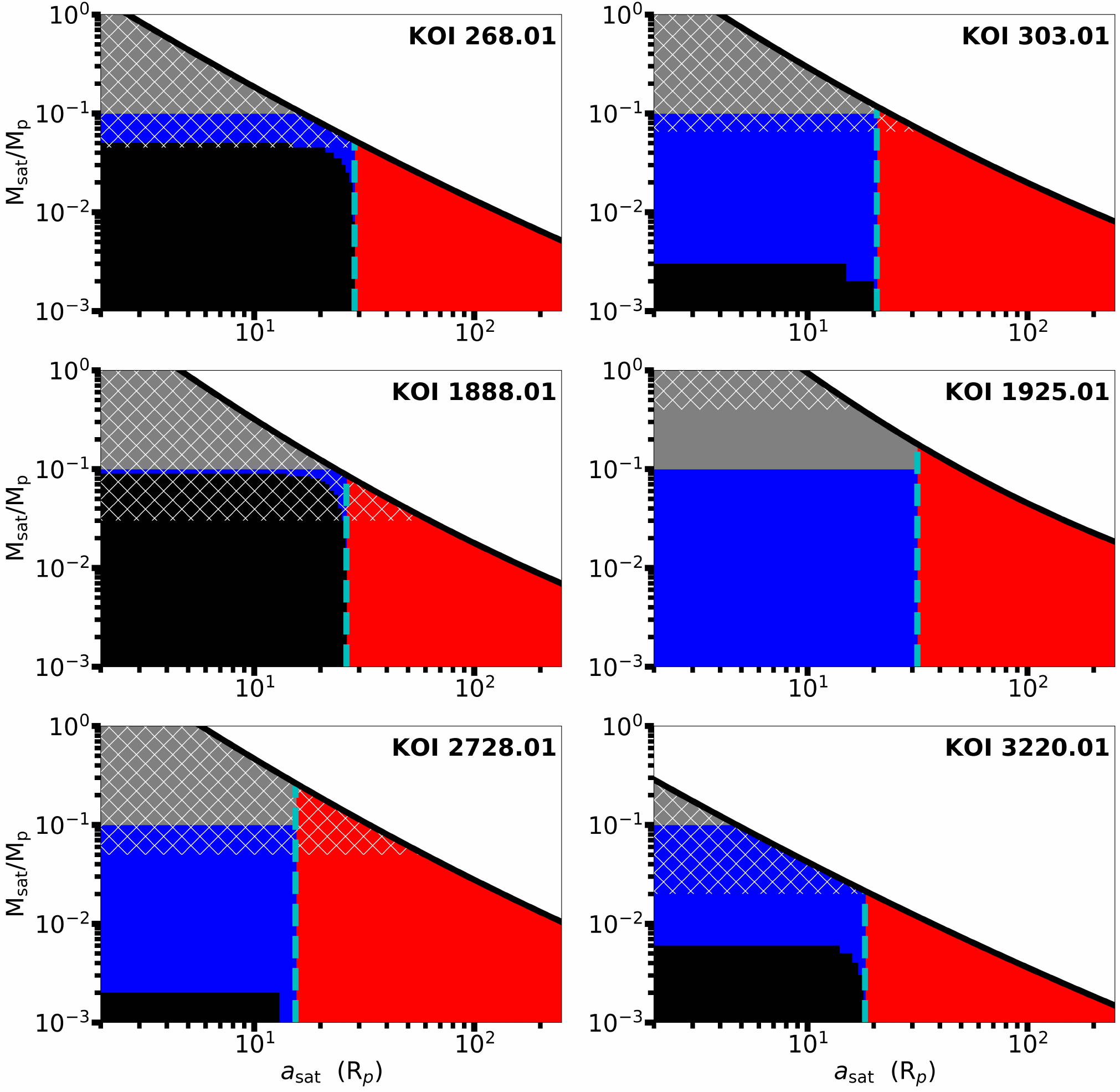}
    \caption{Limits on the planet-satellite mass ratio $M_{\rm sat}/M_{\rm p}$ and satellite separation $a_{\rm sat}$, where regions of parameter space can be excluded based upon orbital stability (red), tidal migration (blue and gray), and observational modeling (white). The black curve marks the 3$\sigma$ upper limits adapted from \cite{Kipping2020b}.  The cyan dashed line delineates the orbital stability boundary.  The hatched (white) regions mark regions that we exclude because the satellite radius $R_m$ is large enough to produce a detectable transit within the \textit{Kepler} data ($R_m \gtrsim 0.5 R_\oplus$) assuming a Mars-like satellite bulk density ($\rho_{\rm sat} = 3.93$ g/cm$^3$). The gray regions mark conditions where the satellite mass becomes significant for the tidal evolution and we evaluate conditions for KOI 1925.01 using our modifications to \cite{Sasaki2012} that allow for larger mass ratios, where this region overlaps with the hatched area for the other KOIs.  The remaining black regions indicate plausible mass-ratios and separations for stable exomoons in these systems. } 
    \label{fig:exomoon_limits}
\end{figure}

Figure \ref{fig:exomoon_limits} shows the combination of constraints as a function of the planet-satellite mass ratio $M_{\rm sat}/M_{\rm p}$ and separation $a_{\rm sat}$ on a logarithmic scale.  The black regions indicate parameters that allow for possibly extant satellites, which remain below the stability limit for at least the system lifetime.  The red and blue regions are excluded based upon orbital stability and tidal migration constraints, respectively.  The tidal migration constraints apply our constraint that $Q_{\rm crit}<200$ for KOI 1925.01 and $Q_{\rm crit}<2000$ for the other KOIs (Figure \ref{fig:exomoon_tides}).  The black curve marks the 3$\sigma$ boundary in TTVs \citep{Kipping2020b} and parameters above the curve (white region) are excluded because the TTV amplitude would be too large.  The gray region represents where the satellite tides could be significant as to prolong the lifetime of the satellite, but in most cases those regions can be excluded because the satellite could produce detectable transits or distortions (hatched white region).  KOI 1925.01 is an exception, but we show in Figure \ref{fig:KOI1925_tides} (cyan and magenta curves) that the combination of stellar tides with the planetary tides causes the satellite to spiral inwards onto its host planet on a timescale less than the system age.  Exomoons in KOI 1925.01 are completely excluded within our parameter space, especially if the planet does indeed have a high eccentricity (Fig. \ref{fig:KOI1925_tides}).  The other KOIs are significantly constrained to less than half of the unconstrained area alone (i.e., below the black curves).

We use the current mean values from the respective parameters in Table \ref{tab:KOI_params}, where the planetary mass and system age are the most uncertain.  The system age affects our calculation of $Q_{\rm crit}$ (Equation \ref{eqn:Q_crit}) linearly and thus the height of the black region in Figure \ref{fig:exomoon_limits} could change by a factor of $\sim$2 if the systems are actually half as old.  Uncertainties in the planetary mass alter the area of the possible moons by a factor of $\sim$4 because of competing dependencies between $a_{\rm crit}$ for orbital stability and $Q_{\rm crit}$ for tidal migration.  Doubling the planetary mass in each case increases the viability of exomoons,  our assumptions on other planetary properties, such as the tidal Love number, should also be updated due to the increased planetary density.  Our results represent a snapshot of the current knowledge without precise planetary masses or eccentricities, where additional observations are needed to produce more accurate results.

\section{Conclusions}\label{sec:conc}
\cite{Kipping2020b} performed an independent analysis of the transit timing variations (TTVs) for the six KOI candidates that \cite{Fox2020} proposed that such TTVs could result from unseen exomoons.  Our study complements the work by \cite{Kipping2020a} by exploring the theoretical constraints for exomoons in these systems based on our previous study for the orbital stability of exomoons \citep{Rosario-Franco2020} and other works that evaluate tidal migration scenarios \citep{Sasaki2012,Chen2017}.  We find that $\sim$50\% of the parameter space can be excluded due to instabilities that occur from orbital stability constraints ($a_{\rm sat} \gtrsim$ 20 R$_{\rm p}$).  Interior to the stability limit, exomoons face additional hurdles due to the tidal migration within the system lifetime.  Four of the KOI candidate systems (KOI 303.01, 1925,01, 2728.01, and 3220.01) are significantly constrained due to tidal migration timescales, where the remaining two systems (KOI 268.01 and 1888.01) could allow for low-mass ($M_{\rm sat}/M_{\rm p} \lesssim 0.03$), close-in exomoons ($a_{\rm sat} \lesssim$ 20 R$_{\rm p}$) exomoons within the current estimates of the system ages.  Observational uncertainty can affect our estimates, where the biggest differences arise through our estimate of the planetary mass M$_{\rm p}$ using a probabilistic framework with \texttt{Forecaster} \citep{Chen2017}.  However, observational constraints due to the TTV amplitude and non-detection of exomoon transits limit the increases to the tidally allowed region due to this uncertainty such that our results remain accurate within a factor of a few.  Our models assume a circular planetary orbit, where relaxing this condition typically halves the extent of exomoon separations due to a much smaller Hill radius at planetary periastron.  Overall, it appears unlikely that the six KOI systems proposed by \cite{Fox2020} can host large enough exomoons to explain the observed TTVs due to a tidal migration constraint on the planet-satellite mass ratio.

Although these six KOIs may not host exomoons, Kepler 1625b-I \citep{Teachey2018a} remains the best exomoon candidate system.  \cite{Rosario-Franco2020} highlighted this assessment in that the host planet orbits much farther from its host star, which diminishes the influence of stellar tides and significantly increases the Hill radius.  Using Equation \ref{eqn:Q_crit}, we find the lower limit for tidal dissipation $Q_{\rm crit} \geq 2000$ for 10 Gyr to be more than sufficient to allow for such a large exomoon.  Kepler 1625b-I is controversial because the data analysis has been contested suggesting that it is an artifact of the data \citep{Kreidberg2019} or due to a blended observation of a planet that is closer to the host star \citep{Heller2019}, but \cite{Teachey2020} show that the exomoon hypothesis is more probable than the other scenarios proposed.  Exomoons, in general, are an evolving prospect where significant care needs to be used while they remain on the bleeding edge of our detection capabilities.

\acknowledgments
M.R.F acknowledges support from the NRAO Gr\"{o}te Reber Fellowship and the Louis Stokes Alliance for Minority Participation Bridge Program at the University of Texas at Arlington.  This research has made use of the NASA Exoplanet Archive, which is operated by the California Institute of Technology, under contract with the National Aeronautics and Space Administration under the Exoplanet Exploration Program.

\facility {Exoplanet Archive}
\software{Forecaster \cite{Chen2017}; scipy \cite{2020SciPy_NMeth}; matplotlib \cite{Hunter2007}}

\bibliography{sample63}{}

\begin{thebibliography}{}
\expandafter\ifx\csname natexlab\endcsname\relax\def\natexlab#1{#1}\fi
\providecommand{\url}[1]{\href{#1}{#1}}
\providecommand{\dodoi}[1]{doi:~\href{http://doi.org/#1}{\nolinkurl{#1}}}
\providecommand{\doeprint}[1]{\href{http://ascl.net/#1}{\nolinkurl{http://ascl.net/#1}}}
\providecommand{\doarXiv}[1]{\href{https://arxiv.org/abs/#1}{\nolinkurl{https://arxiv.org/abs/#1}}}

\bibitem[{{Barnes} \& {O'Brien}(2002)}]{Barnes2002}
{Barnes}, J.~W., \& {O'Brien}, D.~P. 2002, \apj, 575, 1087,
  \dodoi{10.1086/341477}

\bibitem[{{Berger} {et~al.}(2018){Berger}, {Huber}, {Gaidos}, \& {van
  Saders}}]{Berger2018}
{Berger}, T.~A., {Huber}, D., {Gaidos}, E., \& {van Saders}, J.~L. 2018, \apj,
  866, 99, \dodoi{10.3847/1538-4357/aada83}

\bibitem[{{Cabrera} \& {Schneider}(2007)}]{Cabrera2007}
{Cabrera}, J., \& {Schneider}, J. 2007, \aap, 464, 1133,
  \dodoi{10.1051/0004-6361:20066111}

\bibitem[{{Chen} \& {Kipping}(2017)}]{Chen2017}
{Chen}, J., \& {Kipping}, D. 2017, \apj, 834, 17,
  \dodoi{10.3847/1538-4357/834/1/17}

\bibitem[{{Cheng} {et~al.}(2014){Cheng}, {Lee}, \& {Peale}}]{Cheng2014}
{Cheng}, W.~H., {Lee}, M.~H., \& {Peale}, S.~J. 2014, \icarus, 233, 242,
  \dodoi{10.1016/j.icarus.2014.01.046}

\bibitem[{{Eggleton} {et~al.}(1998){Eggleton}, {Kiseleva}, \&
  {Hut}}]{Eggleton1998}
{Eggleton}, P.~P., {Kiseleva}, L.~G., \& {Hut}, P. 1998, \apj, 499, 853,
  \dodoi{10.1086/305670}

\bibitem[{{Fabrycky} \& {Tremaine}(2007)}]{Fabrycky2007}
{Fabrycky}, D., \& {Tremaine}, S. 2007, \apj, 669, 1298, \dodoi{10.1086/521702}

\bibitem[{{Fox} \& {Wiegert}(2020)}]{Fox2020}
{Fox}, C., \& {Wiegert}, P. 2020, arXiv e-prints, arXiv:2006.12997.
\newblock \doarXiv{2006.12997}

\bibitem[{{Gavrilov} \& {Zharkov}(1977)}]{Gavrilov1977}
{Gavrilov}, S.~V., \& {Zharkov}, V.~N. 1977, \icarus, 32, 443,
  \dodoi{10.1016/0019-1035(77)90015-X}

\bibitem[{{Goldreich} \& {Soter}(1966)}]{Goldreich1966}
{Goldreich}, P., \& {Soter}, S. 1966, \icarus, 5, 375,
  \dodoi{10.1016/0019-1035(66)90051-0}

\bibitem[{{Heller}(2014)}]{Heller2014}
{Heller}, R. 2014, \apj, 787, 14, \dodoi{10.1088/0004-637X/787/1/14}

\bibitem[{{Heller}(2018)}]{Heller2018}
---. 2018, \aap, 610, A39, \dodoi{10.1051/0004-6361/201731760}

\bibitem[{{Heller} {et~al.}(2016){Heller}, {Hippke}, \& {Jackson}}]{Heller2016}
{Heller}, R., {Hippke}, M., \& {Jackson}, B. 2016, \apj, 820, 88,
  \dodoi{10.3847/0004-637X/820/2/88}

\bibitem[{{Heller} {et~al.}(2019){Heller}, {Rodenbeck}, \&
  {Bruno}}]{Heller2019}
{Heller}, R., {Rodenbeck}, K., \& {Bruno}, G. 2019, \aap, 624, A95,
  \dodoi{10.1051/0004-6361/201834913}

\bibitem[{{Hippke}(2015)}]{Hippke2015}
{Hippke}, M. 2015, \apj, 806, 51, \dodoi{10.1088/0004-637X/806/1/51}

\bibitem[{Hunter(2007)}]{Hunter2007}
Hunter, J.~D. 2007, Computing in Science \& Engineering, 9, 90,
  \dodoi{10.1109/MCSE.2007.55}

\bibitem[{{Hut}(1981)}]{Hut1981}
{Hut}, P. 1981, \aap, 99, 126

\bibitem[{{Kipping}(2020)}]{Kipping2020b}
{Kipping}, D. 2020, \apjl, 900, L44, \dodoi{10.3847/2041-8213/abafa9}

\bibitem[{{Kipping} \& {Teachey}(2020)}]{Kipping2020a}
{Kipping}, D., \& {Teachey}, A. 2020, arXiv e-prints, arXiv:2004.04230.
\newblock \doarXiv{2004.04230}

\bibitem[{{Kipping}(2009{\natexlab{a}})}]{Kipping2009a}
{Kipping}, D.~M. 2009{\natexlab{a}}, \mnras, 392, 181,
  \dodoi{10.1111/j.1365-2966.2008.13999.x}

\bibitem[{{Kipping}(2009{\natexlab{b}})}]{Kipping2009b}
---. 2009{\natexlab{b}}, \mnras, 396, 1797,
  \dodoi{10.1111/j.1365-2966.2009.14869.x}

\bibitem[{{Kipping} {et~al.}(2012){Kipping}, {Bakos}, {Buchhave},
  {Nesvorn{\'y}}, \& {Schmitt}}]{Kipping2012}
{Kipping}, D.~M., {Bakos}, G.~{\'A}., {Buchhave}, L., {Nesvorn{\'y}}, D., \&
  {Schmitt}, A. 2012, \apj, 750, 115, \dodoi{10.1088/0004-637X/750/2/115}

\bibitem[{{Kipping} {et~al.}(2013{\natexlab{a}}){Kipping}, {Forgan}, {Hartman},
  {Nesvorn{\'y}}, {Bakos}, {Schmitt}, \& {Buchhave}}]{Kipping2013b}
{Kipping}, D.~M., {Forgan}, D., {Hartman}, J., {et~al.} 2013{\natexlab{a}},
  \apj, 777, 134, \dodoi{10.1088/0004-637X/777/2/134}

\bibitem[{{Kipping} {et~al.}(2013{\natexlab{b}}){Kipping}, {Hartman},
  {Buchhave}, {Schmitt}, {Bakos}, \& {Nesvorn{\'y}}}]{Kipping2013a}
{Kipping}, D.~M., {Hartman}, J., {Buchhave}, L.~A., {et~al.}
  2013{\natexlab{b}}, \apj, 770, 101, \dodoi{10.1088/0004-637X/770/2/101}

\bibitem[{{Kipping} {et~al.}(2014){Kipping}, {Nesvorn{\'y}}, {Buchhave},
  {Hartman}, {Bakos}, \& {Schmitt}}]{Kipping2014}
{Kipping}, D.~M., {Nesvorn{\'y}}, D., {Buchhave}, L.~A., {et~al.} 2014, \apj,
  784, 28, \dodoi{10.1088/0004-637X/784/1/28}

\bibitem[{{Kipping} {et~al.}(2015){Kipping}, {Schmitt}, {Huang}, {Torres},
  {Nesvorn{\'y}}, {Buchhave}, {Hartman}, \& {Bakos}}]{Kipping2015b}
{Kipping}, D.~M., {Schmitt}, A.~R., {Huang}, X., {et~al.} 2015, \apj, 813, 14,
  \dodoi{10.1088/0004-637X/813/1/14}

\bibitem[{{Kreidberg} {et~al.}(2019){Kreidberg}, {Luger}, \&
  {Bedell}}]{Kreidberg2019}
{Kreidberg}, L., {Luger}, R., \& {Bedell}, M. 2019, \apjl, 877, L15,
  \dodoi{10.3847/2041-8213/ab20c8}

\bibitem[{{Lainey}(2016)}]{Lainey2016}
{Lainey}, V. 2016, Celestial Mechanics and Dynamical Astronomy, 126, 145,
  \dodoi{10.1007/s10569-016-9695-y}

\bibitem[{{Lewis} {et~al.}(2015){Lewis}, {Ochiai}, {Nagasawa}, \&
  {Ida}}]{Lewis2015}
{Lewis}, K.~M., {Ochiai}, H., {Nagasawa}, M., \& {Ida}, S. 2015, \apj, 805, 27,
  \dodoi{10.1088/0004-637X/805/1/27}

\bibitem[{{Morton} {et~al.}(2016){Morton}, {Bryson}, {Coughlin}, {Rowe},
  {Ravichandran}, {Petigura}, {Haas}, \& {Batalha}}]{Morton2016}
{Morton}, T.~D., {Bryson}, S.~T., {Coughlin}, J.~L., {et~al.} 2016, \apj, 822,
  86, \dodoi{10.3847/0004-637X/822/2/86}

\bibitem[{{Rogers}(2015)}]{Rogers2015}
{Rogers}, L.~A. 2015, \apj, 801, 41, \dodoi{10.1088/0004-637X/801/1/41}

\bibitem[{{Rosario-Franco} {et~al.}(2020){Rosario-Franco}, {Quarles},
  {Musielak}, \& {Cuntz}}]{Rosario-Franco2020}
{Rosario-Franco}, M., {Quarles}, B., {Musielak}, Z.~E., \& {Cuntz}, M. 2020,
  \aj, 159, 260, \dodoi{10.3847/1538-3881/ab89a7}

\bibitem[{{Sartoretti} \& {Schneider}(1999)}]{Sartoretti1999}
{Sartoretti}, P., \& {Schneider}, J. 1999, \aaps, 134, 553,
  \dodoi{10.1051/aas:1999148}

\bibitem[{{Sasaki} {et~al.}(2012){Sasaki}, {Barnes}, \& {O'Brien}}]{Sasaki2012}
{Sasaki}, T., {Barnes}, J.~W., \& {O'Brien}, D.~P. 2012, \apj, 754, 51,
  \dodoi{10.1088/0004-637X/754/1/51}

\bibitem[{{Silva Aguirre} {et~al.}(2015){Silva Aguirre}, {Davies}, {Basu},
  {Christensen-Dalsgaard}, {Creevey}, {Metcalfe}, {Bedding}, {Casagrande},
  {Handberg}, {Lund}, {Nissen}, {Chaplin}, {Huber}, {Serenelli}, {Stello}, {Van
  Eylen}, {Campante}, {Elsworth}, {Gilliland}, {Hekker}, {Karoff}, {Kawaler},
  {Kjeldsen}, \& {Lundkvist}}]{SilvaAguirre2015}
{Silva Aguirre}, V., {Davies}, G.~R., {Basu}, S., {et~al.} 2015, \mnras, 452,
  2127, \dodoi{10.1093/mnras/stv1388}

\bibitem[{{Sucerquia} {et~al.}(2019){Sucerquia}, {Alvarado-Montes}, {Zuluaga},
  {Cuello}, \& {Giuppone}}]{Sucerquia2019}
{Sucerquia}, M., {Alvarado-Montes}, J.~A., {Zuluaga}, J.~I., {Cuello}, N., \&
  {Giuppone}, C. 2019, \mnras, 489, 2313, \dodoi{10.1093/mnras/stz2110}

\bibitem[{{Teachey} {et~al.}(2020){Teachey}, {Kipping}, {Burke}, {Angus}, \&
  {Howard}}]{Teachey2020}
{Teachey}, A., {Kipping}, D., {Burke}, C.~J., {Angus}, R., \& {Howard}, A.~W.
  2020, \aj, 159, 142, \dodoi{10.3847/1538-3881/ab7001}

\bibitem[{{Teachey} \& {Kipping}(2018)}]{Teachey2018b}
{Teachey}, A., \& {Kipping}, D.~M. 2018, Science Advances, 4, eaav1784,
  \dodoi{10.1126/sciadv.aav1784}

\bibitem[{{Teachey} {et~al.}(2018){Teachey}, {Kipping}, \&
  {Schmitt}}]{Teachey2018a}
{Teachey}, A., {Kipping}, D.~M., \& {Schmitt}, A.~R. 2018, \aj, 155, 36,
  \dodoi{10.3847/1538-3881/aa93f2}

\bibitem[{{Tokadjian} \& {Piro}(2020)}]{Tokadjian2020}
{Tokadjian}, A., \& {Piro}, A.~L. 2020, arXiv e-prints, arXiv:2007.01487.
\newblock \doarXiv{2007.01487}

\bibitem[{{Virtanen} {et~al.}(2020){Virtanen}, {Gommers}, {Oliphant},
  {Haberland}, {Reddy}, {Cournapeau}, {Burovski}, {Peterson}, {Weckesser},
  {Bright}, {van der Walt}, {Brett}, {Wilson}, {Jarrod Millman}, {Mayorov},
  {Nelson}, {Jones}, {Kern}, {Larson}, {Carey}, {Polat}, {Feng}, {Moore}, {Vand
  erPlas}, {Laxalde}, {Perktold}, {Cimrman}, {Henriksen}, {Quintero}, {Harris},
  {Archibald}, {Ribeiro}, {Pedregosa}, {van Mulbregt}, \&
  {Contributors}}]{2020SciPy_NMeth}
{Virtanen}, P., {Gommers}, R., {Oliphant}, T.~E., {et~al.} 2020, Nature
  Methods, 17, 261, \dodoi{https://doi.org/10.1038/s41592-019-0686-2}

\bibitem[{{Ward} \& {Reid}(1973)}]{Ward1973}
{Ward}, W.~R., \& {Reid}, M.~J. 1973, \mnras, 164, 21,
  \dodoi{10.1093/mnras/164.1.21}

\end{thebibliography}
\bibliographystyle{aasjournal}

\clearpage

\begin{deluxetable}{lcccccc}
\tablecolumns{7}
\tablecaption{Parameters for the 6 exomoon candidate KOIs. \label{tab:KOI_params}}
\tablehead{\colhead{KOI} & \colhead{M$_\star$} & \colhead{$R_{\rm p}$} & \colhead{M$_{\rm p}$\tablenotemark{$\dagger$}}& \colhead{$a_{\rm p}$} & \colhead{$\tau$} & \colhead{References} \\
\colhead{} & \colhead{(M$_\odot$)} & \colhead{($R_\oplus$)} & \colhead{(M$_\oplus$)} & \colhead{(AU)} & \colhead{(Gyr)} &}
\startdata
268.01  & 1.175$^{+0.058}_{-0.064}$ & 3.32$^{+0.85}_{-0.64}$ & 10.4$^{+11.1}_{-5.5}$ & 0.4756 & 3.05$^{+0.85}_{-0.64}$ & a,b\\
303.01  & 0.871$^{+0.142}_{-0.142}$ & 2.78$^{+0.39}_{-0.38}$ & 8.13$^{+6.70}_{-3.67}$ & 0.2897 & 6.31$^{+3.15}_{-3.81}$ & a,c\\
1888.01 & 1.406$^{+0.086}_{-0.086}$ & 4.76$^{+0.34}_{-0.31}$ & 18.6$^{+16.4}_{-8.4}$  & 0.5337 & 1.26$^{+0.33}_{-0.18}$ & a,b\\
1925.01 & 0.890$^{+0.009}_{-0.011}$ & 1.10$^{+0.05}_{-0.04}$ & 1.37$^{+0.88}_{-0.44}$ & 0.3183 & 6.98$^{+0.4}_{-0.5}$ & a,b\\
2728.01 & 1.450$^{+0.601}_{-0.271}$ & 3.224$^{+0.213}_{-0.159}$ & 10.4$^{+9.00}_{-4.71}$ & 0.2743 & 1.700$^{+0.530}_{-0.392}$ & a,b,d\\
3220.01 & 1.340$^{+0.054}_{-0.051}$ & 5.559$^{+0.252}_{-0.889}$ & 25.2$^{+24.2}_{-12.6}$ & 0.4039 & 1.700$^{+0.556}_{-0.459}$ & a,b,d\\
\enddata
\tablenotetext{a}{Kepler Exoplanet Archive DR25}
\tablenotetext{b}{\cite{SilvaAguirre2015}}
\tablenotetext{c}{\cite{Morton2016}}
\tablenotetext{d}{\cite{Berger2018}}
\tablenotetext{\dagger}{Planet masses M$_{\rm p}$ are estimated probabilistically using the planet radius $R_{\rm p}$ \citep{Chen2017}.}
\end{deluxetable}

\end{document}